\documentclass[twocolumn,showpacs,preprintnumbers,amsmath,amssymb,superscriptaddress]{revtex4-1}

\usepackage{graphicx}
\usepackage{dcolumn}
\usepackage{bm}
\usepackage{xcolor}

\begin{document}


\title{Probing quantum criticality in ferromagnetic CeRh\(_6\)Ge\(_4\)}

\author{S.M. Thomas}
\email{smthomas@lanl.gov}
\affiliation{MPA-Quantum, Los Alamos National Laboratory, Los Alamos, NM 87544 USA}
\author{S. Seo}
\affiliation{MPA-Quantum, Los Alamos National Laboratory, Los Alamos, NM 87544 USA}
\affiliation{Department of Physics, Changwon National University, Changwon 51140, South Korea}
\author{T. Asaba}
\author{F. Ronning}
\author{P.F.S. Rosa}
\author{E.D. Bauer}
\author{J.D. Thompson}
\affiliation{MPA-Quantum, Los Alamos National Laboratory, Los Alamos, NM 87544 USA}

\date{\today}

\begin{abstract}
CeRh\(_6\)Ge\(_4\) is unusual in that its ferromagnetic transition can be suppressed continuously to zero temperature, \textit{i.e.}, to a ferromagnetic quantum-critical point (QCP), through the application of modest hydrostatic pressure. This discovery has raised the possibility  that the ferromagnetic QCP may be of the Kondo-breakdown type characterized by a jump in Fermi volume, to which thermopower S measurements should be sensitive. Though $S/T$ changes both sign and magnitude around the critical pressure P$_{c}\approx{}0.8$~GPa, 
these changes are not abrupt but extend over a pressure interval from within the ferromagnetic state up to P$_c$.  
Together with temperature and pressure variations in electrical resistivity and previously reported heat capacity, thermopower results point to the near coincidence of two sequential effects near P$_c$, delocalization of 4f degrees-of-freedom through orbital-selective hybridization followed by quantum criticality of itinerant ferromagnetism. 
\end{abstract}

\maketitle

Clean ferromagnetic (FM) materials typically avoid a quantum-critical point (QCP) by triggering either a transition to a different phase at finite temperature or by undergoing a first-order transition~\cite{Brando2016,Chen2022}.
Nevertheless,  quantum criticality in a system of ferromagnetic itinerant electrons is possible theoretically if the material is noncentrosymmetric with strong spin-orbit coupling, is quasi-one-dimensional, or is sufficiently dirty, \textit{i.e.}, has a short electronic mean-free path~\cite{Kirkpatrick2020}.
Alternatively, if the magnetic electrons are localized, a continuous FM quantum-phase transition can be foreseen within the framework of local, Kondo-breakdown criticality.
In this scenario Kondo coupling of localized and conduction-electron  moments above $P_c$ is suppressed concurrently with the development of ferromagnetic order and produces a jump in the Fermi volume from `large' above $P_c$ to `small' below P$_c$~\cite{Shen2020,Komijani2018}.
This is the scenario proposed for CeRh$_6$Ge$_4$, a rare example of ferromagnetic quantum criticality in which its Curie temperature can be suppressed continuously from 2.5 K at atmospheric pressure to zero temperature $via$ hydrostatic pressure without the emergence of an intervening phase transition below the critical pressure P$_{c}\approx{}0.8$~GPa~\cite{Shen2020, Kotegawa2019}.
Support for Kondo-breakdown criticality has come primarily from a comparison of de Haas-van Alphen (dHvA) measurements at atmospheric pressure with spin-polarized density-function calculations that include spin-orbit coupling and that assume the magnetic 4f electrons of Ce are either localized or itinerant, a comparison that is argued to be consistent with the 4$f$ electrons in CeRh\(_6\)Ge\(_4\) being localized at atmospheric pressure~\cite{Wang2021}.
Further, the temperature dependence of dHvA oscillations show that k$_F$l$_{tr}$ $>$ 300, where k$_F$ is the Fermi momentum and l$_{tr}$ is the electronic mean free path.
Such a large k$_F$l$_{tr}$ implies that the crystal is solidly in the clean limit, \textit{i.e.}, quench disorder is irrelevant to the observation of  FM criticality.

Other experiments, however, raise questions about the localized nature of the 4f electrons.
At atmospheric pressure, the specific heat divided by temperature (C/T) remains quite large at milliKelvin temperatures deep in the ordered state ($\approx{}400$ mJ~mol$^{-1}$~K$^{-2}$), entropy recovered up to $T_C$ is only 0.19~$R\ln{2}$, and the ordered moment is 0.28~$\mu{}_B/Ce$, which is well below that expected (1.28~$\mu{}_B/Ce$) for localized $4f$ electrons in a $\Gamma_7$ crystal-electric field (CEF) doublet ground state~\cite{Wu2021,Shen2020}.
Each of these is consistent with Kondo hybridization in which the spin of the local moment becomes part of the Fermi volume, creating a `large' Fermi surface.
Though zero-point fluctuations of local moments could play a role by mimicking expectations of Kondo hybridization in thermodynamic measurements~\cite{Shen2020}, angle-resolved photoemission spectroscopy (ARPES) reveals considerable anisotropic hybridization between $4f$ and conduction ($c$) electrons~\cite{Wu2021}, which also is implied by optical spectroscopy~\cite{Pei2021a}.

Real-space electron densities are consistent with anisotropic hybridization observed in ARPES~\cite{Wu2021}. Wavefunctions of both the CEF ground state doublet and the $\Gamma_9$ first excited doublet at $5.8$~meV display electron density primarily out of the hexagonal basal plane, in agreement with strong $c$-axis hybridization and easy-plane magnetic anisotropy in CeRh$_6$Ge$_4$.
The $\Gamma_9$ doublet, whose wavefunction has greater spatial extent perpendicular to the $c$-axis, is argued to hybridize even more strongly with conduction-band states, an indication of a stronger Kondo coupling compared to the ground state.
Such CEF properties may explain the discrepancy between the observation of anisotropic Kondo hybridization and the localized $4f$ character inferred from quantum oscillations at atmospheric pressure~\cite{Shu2021}.
Recent dHvA measurements as a function of pressure, however, find that cyclotron frequencies, a measure of extremal orbits on the Fermi surface, are unchanged from below to above P$_c$~\cite{Huiqiu2023} and may further question a Kondo-breakdown scenario.
Nevertheless, these measurements, performed in high magnetic fields, so-far only detect electron masses m* of less than 10 times the mass of free electrons (m$_e$), which would seem not to account for the large $C/T$ in this material and leaves open the possibility of changes in cyclotron frequencies of higher mass orbits at P$_c$.

With a non-centrosymmetric crystal structure and a chain-structure of Ce atoms along the hexagonal c-axis of CeRh\(_6\)Ge\(_4\), existing results are enigmatic---some pointing to the possibility of itinerant-type ferromagnetic criticality and others to a local-moment (Kondo-breakdown) scenario.
Thermodynamic and electrical transport measurements, though clearly signaling non-Fermi-liquid characteristics of quantum criticality around the critical pressure of CeRh$_6$Ge$_4$, are unable to distinguish between these two possibilities.
Experiments that directly probe the Fermi-surface as a function of pressure without the need for large applied magnetic fields could be beneficial in helping resolve the conundrum.
Indeed, thermopower measurements have been effective in revealing the nature of field-tuned quantum criticality in YbRh$_2$Si$_2$ and pressure-tuned criticality in CeRh$_{0.58}$Ir$_{0.42}$In$_5$~\cite{Hartmann2010, Luo2018}.

Here, we report measurements of the pressure-dependent thermopower ($S$) and electrical resistivity $\rho$ of CeRh$_6$Ge$_4$.
At low pressure, the magnitude of $S/T$ in the low-temperature limit remains nearly constant, which suggests that any FS changes there are minor.
At $\sim$$0.7$~GPa, however, $S/T$ changes sign and increases smoothly through the magnetic QCP at P$_{c}\approx{}0.8$~GPa before saturating to a larger value at higher pressures.
Field-dependent measurements of $S/T$ argue for the increase at high pressures being an intrinsic response to a change in Fermi surface and not to the loss of an internal magnetic field accompanying ferromagnetic order.

\begin{figure}
	\begin{center}
		\includegraphics[width=0.92\columnwidth]{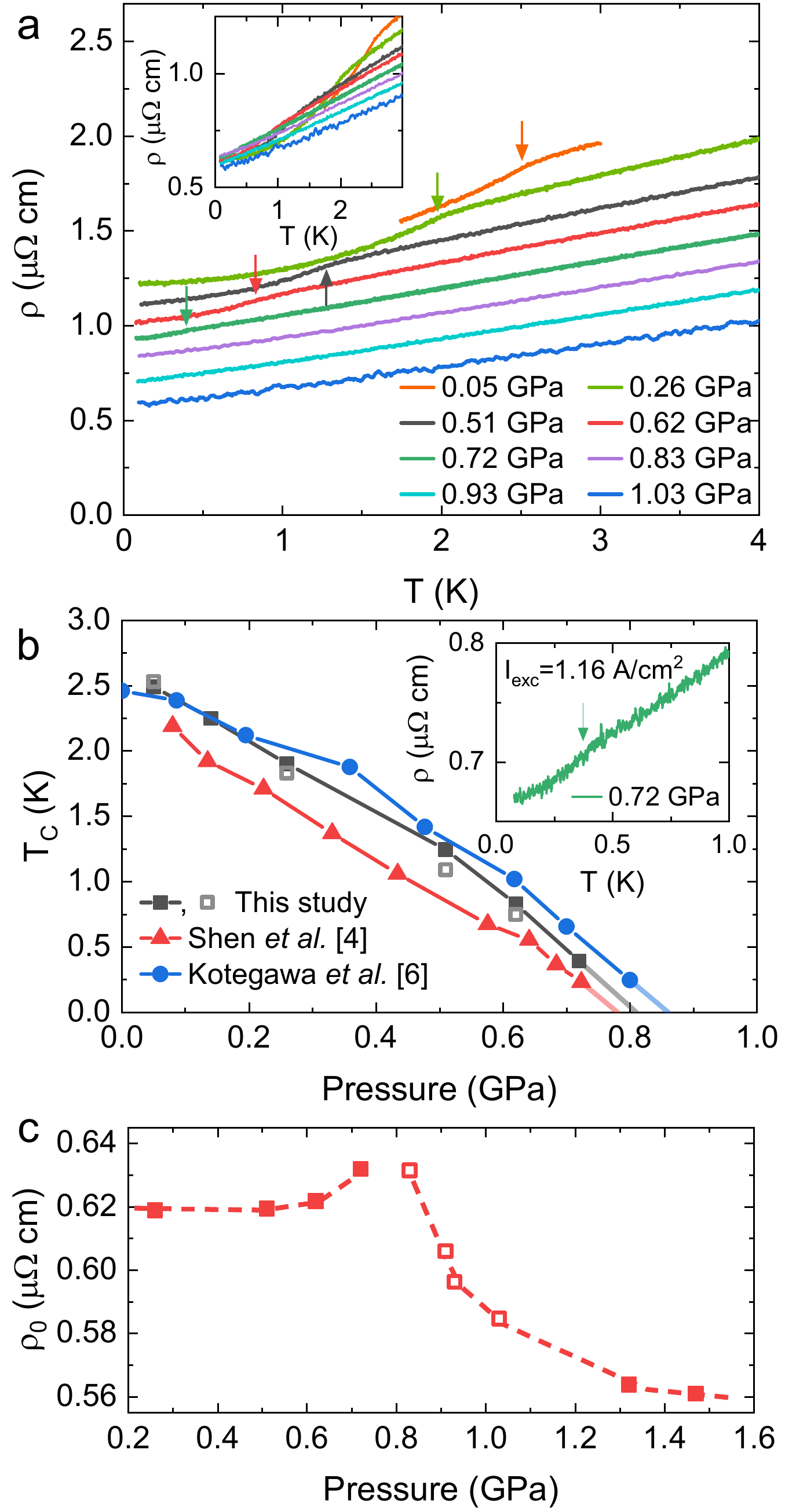}
		\caption{
			(a)~Low-temperature resistivity versus temperature at a number of different pressures. The data is offset for clarity. The inset shows the same data without an offset.
            (b)~Pressure-temperature phase diagram obtained in this study compared with other published studies.
			Solid squares are determined from resistivity, whereas open squares are determined from thermopower.
            The inset shows T$_C$ determined by a change in slope of the electrical resistivity (marked by an arrow) at 0.72 GPa.
			(c)~Residual resistivity from a fit to $\rho{}\left(T\right)=AT^2+\rho_0$ (closed symbols) or $\rho{}\left(T\right)=A'T+\rho_0$ (open symbols).
			Uncertainty in the fitted value of $\rho_0$ is smaller than the size of the data markers.
		}\label{fig:fig1}
	\end{center}
\end{figure}

Single crystals of CeRh\(_6\)Ge\(_4\) were grown using the Bi-flux technique~\cite{Vosswinkel}.
X-ray diffraction confirmed the P$\overline{6}$m2 hexagonal structure in which chains of Ce atoms form along the $c$-axis with a Ce-Ce spacing about half that in the perpendicular direction. 
Thermopower measurements were performed using a steady-state technique~\cite{Luo2016}.
One end of the sample was attached to a heater and the other end was thermally anchored.
A pair of Chromel-Au\(_{99.93\%}\)Fe\(_{0.07\%}\) thermocouples was calibrated as a function of magnetic field and temperature~\cite{Stockert2011} and used to measure the temperature gradient ($\Delta{}T$).
A pair of spot-welded voltage contacts in line with the thermocouples was used to measure the voltage ($E_x$).
The same contacts were used with additional current leads to measure electrical resistivity, $\rho_{xx}$, in a standard 4-point configuration.
Heat and electrical current were applied parallel to the \textit{c}-axis. In thermopower measurements, the average temperature of the sample was determined by adding an offset to the temperature measured by a primary thermometer attached to the pressure cell.
This offset was determined by measuring the change in resistivity of the sample before and after applying heat using the same thermal profile as during the thermopower measurement.
Using this approach, the average sample temperature was determined to be approximately $T_0+3\Delta{}T$, where $T_0$ is the base temperature and $\Delta{}T$ (typically 20--30 mK) was the temperature gradient measured by the thermocouple.
The thermopower is defined as $S_{xx}=-E_x/\Delta{}T$.
All measurements were performed in a piston-clamp pressure cell using Daphne oil as a hydrostatic pressure medium.
A lead manometer was used to determine the pressure.

Figure~1a shows the temperature-dependent electrical resistivity as a function of pressure.
For clarity, an offset 0.1~$\mu\Omega$~cm is added to each curve.
In the non-offset data, shown in the inset of Fig.~1a, the residual resistivity extrapolates to $0.62~\mu{}\Omega{}$~cm at 0.05 GPa giving a residual resistivity ratio (RRR $=\rho_{300K}/\rho_{0}$) of $48$.
The residual resistivity ratio here is similar to the value of 45 reported by Shen \textit{et al.}~\cite{Shen2020} and larger than the value of 30 reported in Kotegawa \textit{et al.}~\cite{Kotegawa2019}.
A kink in $\rho${(T,P)} indicates the onset of FM order (T$_C$, marked with arrows) at low pressures.
This anomaly is still clearly observed at 0.72~GPa (see inset of Fig.~1b), which is consistent with prior specific heat results~\cite{Shen2020}.
From these data, we determine the \textit{T$_C$-P} phase diagram in Fig.~1b.
A linear fit of the last three points in the diagram extrapolates to a critical pressure of P$_{c} \sim 0.8$~GPa that is between values obtained in prior reports~\cite{Shen2020,Kotegawa2019}.
At this pressure, the extrapolated residual resistivity diverges (Fig. 1c) as expected due to a renormalization of the impurity scattering potential at a ferromagnetic QCP~\cite{Miyake2002}.

We turn to thermopower  measurements under pressure.
Figure~2a shows $S/T$  for pressures less than P$_c$, whereas data for pressures greater than P$_c$ are shown in Fig.~2b.
As denoted by arrows in Fig. 2a, $S/T$ is sensitive to the onset of magnetic order, increasing below T$_C$.
These temperatures are included as gray open squares in Fig. 1b.
Though clear in resistivity measurements, there is not an obvious signature for magnetic order in $S/T$ above 200 mK at 0.72 GPa, but T$_C$ at this pressure is close to the lowest temperature at which $S/T$ is measured, which could make detecting T$_C$ difficult. 
Nevertheless, a comparison of data in Figs. 2a and b  shows that there is a pronounced increase in the magnitude of $S/T$  at lowest temperatures as pressure increases above 0.62~GPa.
This is obvious in Fig. 2c where we plot $S/T$ at 200 mK as a function of pressure.
Within error bars, $S/T$ is essentially constant up to 0.62 GPa above which it begins to increase and changes sign inside the magnetically ordered phase before plateauing at higher pressures.
The marked increase in $S/T$ that extends from below to P$_c$ is not due to a loss of internal magnetic field arising from ferromagnetic order, which terminates at P$_c$.
Substantiation of this conclusion is demonstrated in Fig.~3a where we see in the paramagnetic state at low temperatures that $S/T$ is suppressed by an externally applied field, contrary to the increase in $S/T$ when the system orders (Fig.~2a) and produces a net internal magnetic field. 

Thermopower is  highly sensitive to Fermi-surface changes~\cite{Behnia2004}, and the pronounced pressure variation of $S/T$ around P$_c$ strongly suggests changes in the Fermi surface, which might support a Kondo-breakdown scenario of the quantum criticality.
In a generalization of this scenario~\cite{Vojta2010}, theory predicts that, independent of the nature of the magnetic order and at sufficiently low temperature, there should be a sharp feature in the magnitude of $S/T$ at the QCP below an energy scale $E^*\approx0.1{\left(q^*/k_F^c\right)}^3T_0$~\cite{Kim2010}.
Here, $q^*$ is the difference in wavevector between  conduction and spinon Fermi surfaces, $k_F^c$ is the Fermi wavevector of the conduction electrons, and $T_0$ is the temperature scale at which $R\ln{2}$ entropy is recovered.
Up to 5~K, CeRh$_6$Ge$_4$ only recovers approximately 0.3 of $R\ln{2}$, but an extrapolation of the low-temperature specific heat to higher temperatures suggests that $T_0$ is on the order of 20~K, estimated by assuming a spin-1/2 Kondo model~\cite{Matsuoka2015}.
Because the ratio $q^*/k_F^c$ is at most unity and likely much smaller, $E^*$ may be at an inaccessibly low temperature.
The absence of the theoretically predicted feature in $S/T$ at temperatures above 200 mK does not support but also does not rule out a Kondo-breakdown scenario of criticality.

\begin{figure*}
	\begin{center}
		\includegraphics[width=1\textwidth]{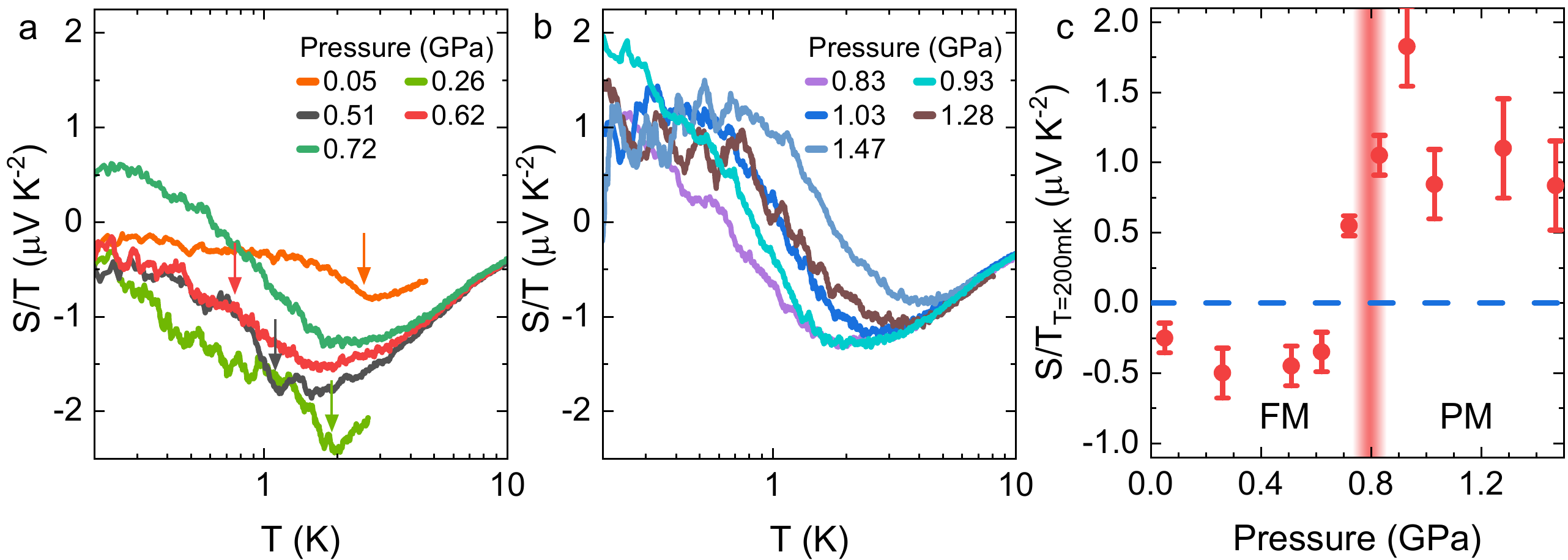}
		\caption{
			(a)~Thermopower divided by temperature versus temperature for pressures less than P$_c$.
			The arrows indicate T$_C$.
			(b)~Thermopower divided by temperature versus temperature for pressures greater than P$_c$.
            (c)~Thermopower divided by temperature at 200~mK versus pressure. The dashed blue line corresponds to a value of zero.
		}\label{fig:fig2}
	\end{center}
\end{figure*}

In a simple free-electron model, thermopower probes the specific heat per electron, which allows the definition of a dimensionless quantity q that is equal to the number of carriers per formula unit~\cite{Behnia2004}:
\begin{equation}
	q=\frac{S}{T}\Bigr|_{T=0}\frac{N_{AV}e}{\gamma},
\end{equation}
where $\gamma$ is the Sommerfeld coefficient, $N_{AV}$ is Avogadro's number, and $e$ is the elementary charge.
Because both specific heat and thermopower are dominated by  bands with the heaviest masses at low temperature~\cite{Miyake2005}, q is of order unity in several heavy-fermion Ce compounds~\cite{Behnia2004}.
A significant departure from unity may indicate either a carrier density of less than one per formula unit,  compensation between electron and hole bands with similar effective mass, or an enhanced $\gamma$ from zero-point fluctuations.

Using a value of $\gamma$ of approximately 0.4 J~mol$^{-1}$~K$^{-2}$~\cite{Shen2020} and $S/T$ of $-0.25$ $\mu$V K$^{-2}$, q at P $\approx{}0$ is only -0.06, which is in stark contrast to the expected value of order unity.
At the highest pressure (1.12~GPa) for which heat capacity data are available~\cite{Shen2020} for a comparison to $S/T$, q changes sign and  rises to +0.18, still far from unity.
Quantum-oscillation experiments indicate the presence of both hole and electron pockets, in agreement with density function theory calculations~\cite{Wang2021}.
This is  likely the primary contributor to the reduced value of q, but we cannot rule out less likely possibilities of a low carrier density or zero-point fluctuations.
Considering the existence of both electron and hole pockets, the positive sign of $S/T$ at pressures above 0.7 GPa indicates that a contribution from the hole pocket becomes more pronounced at high pressures.
This, however, is seemingly inconsistent with a report of the Hall coefficient ($R_H$) that becomes more negative, \textit{i.e.}, in a simple picture, either a higher density of electrons or a smaller density of holes at pressures above P$_c$ than at P=0~\cite{Huiqiu2023}. 
It is not immediately obvious how to reconcile different conclusions from thermopower and Hall measurements, though thermopower above 0.7 GPa is weighted by  massive charge (hole) carriers and lighter, more mobile electron carriers could increasingly dominate $R_H$. Nevertheless, Hall measurements find that $d(-R_H)/dP$ increases near 0.6 GPa which, like the initial increase in $S/T$ (Fig.~2c), is within the ordered phase.
In spite of lacking a definitive explanation for the origin of a reduced value of q, changes in the sign and magnitude of $S/T$ are unambiguous.
The important observation is that these changes occur within the magnetically ordered state at $\sim0.7$~GPa and strongly suggest a Fermi-surface change in that regime.  

The results of Fig. 2c suggest two sequential effects: a change in Fermi surface followed at higher pressures by a quantum-critical point at P$_c$.
One possible interpretation of these observations is provided by the multipolar Bose-Fermi Kondo model in which two sequential QCPs are expected, one in spin degrees-of-freedom and the other in the orbital channel, even though both degrees-of-freedom are coupled by the spin-orbit interaction~\cite{Liu2023a}.
Both are Kondo-breakdown-type QCPs with an associated increase in Fermi volume from small to large.
This model has been used successfully to account for sequential QCPs in Ce$_3$Pd$_{20}$Si$_6$~\cite{Martelli2019}.
A condition for the applicability of this model is that the orbital channel is relevant.

As mentioned in the introduction, inelastic neutron scattering and ARPES find that hybridization is strong and anisotropic.
Significantly, there are no well-defined crystal-field excitations in the neutron spectrum at energy transfers to 80 meV; instead, there is broad magnetic scattering from less than 1.5 meV to at least 60 meV.
Nevertheless, it is possible to account for this magnetic scattering by assuming quasielastic scattering in the CEF ground state with a full width at half maximum  (FWHM) of about 3.3 meV and a much broader inelastic excitation with FWHM of about 30 meV~\cite{Shu2021}.
With ground and first excited CEF doublets separated by about 5.8 meV, estimated from a CEF analysis of anisotropic magnetic susceptibility, these neutron scattering results clearly imply a mixing of ground and first excited CEF wavefunctions to create an effective four-fold degenerate ground state.
This conclusion is supported both by the strong $f-c$ hybridization detected in ARPES~\cite{Wu2021} and by a Kadowaki-Woods ratio that corresponds to a ground state degeneracy of 4~\cite{Shen2020, Wu2021}.
Consequently, orbital degrees-of-freedom are relevant, and the multiorbital Bose-Fermi Kondo model might account for two nearby QCPs that involve a change of Fermi surface implied by our thermopower results.
We would expect, however, to find evidence for two sequential jumps in Fermi volume in the pressure range $\approx{}0.7$ GPa to P$_c$ and for evidence of orbital order.
These expectations are not obvious in our data, but the close proximity of two QCPs might create just a broadened response to pressures where the Fermi surface sequentially reconstructs.
This possibility is questioned by isothermal plots of S/T versus pressure shown in Fig.~3b where we see that the response does not change noticeably when temperature is reduced from 400 to 200 mK, a 50\% change.

Taking orbital degrees-of-freedom to be relevant, there is an alternative interpretation, namely pressure-dependent orbital-selective  hybridization, that is suggested by ARPES which finds two symmetry-inequivalent 4$f$ bands with different orbital characters at the Fermi energy~\cite{Wu2021}.
One band has notably weaker spectral weight which  implies weaker $f-c$ hybridization.
Spectral weight in this band decreases more rapidly with increasing temperature than that in the more strongly hybridized band, and, we associate it with the CEF ground state doublet that also has some orbital character of the $\Gamma_9$ excited CEF doublet.
Applied pressure increases $f-c$ hybridization in Ce-based compounds in which the relative rate of increase in hybridization is larger for states with the lower characteristic (Kondo) energy scale~\cite{Thompson1994}.
From neutron scattering, this energy scale in CeRh$_6$Ge$_4$ is roughly an order of magnitude smaller in the CEF ground state (with primarily $\Gamma_7$ character) than in the first excited CEF state (with primarily $\Gamma_9$ character)~\cite{Shu2021}. 
We, therefore, expect pressure to preferentially increase hybridization of states with primarily $\Gamma_7$ character relative to those of primarily $\Gamma_9$ character.
This will mix even more of the excited CEF into the ground state and, with the $\Gamma_9$ wavefunction tending to hybridize more readily with $c$-states, will increase the Fermi volume already enlarged relative to the limit that the 4$f$ electrons are completely localized.
Data in Fig.~2c show this change in Fermi surface begins somewhat below 0.7 GPa.
We note that the 4$f$ band with more spectral weight at $E_F$, which we identify as having primarily $\Gamma_9$ character, develops from 4$f$ hybridization with a hole band.
It is not surprising then that $S/T$ changes from negative to positive with increasing pressure.

\begin{figure}
	\begin{center}
		\includegraphics[width=0.88\columnwidth]{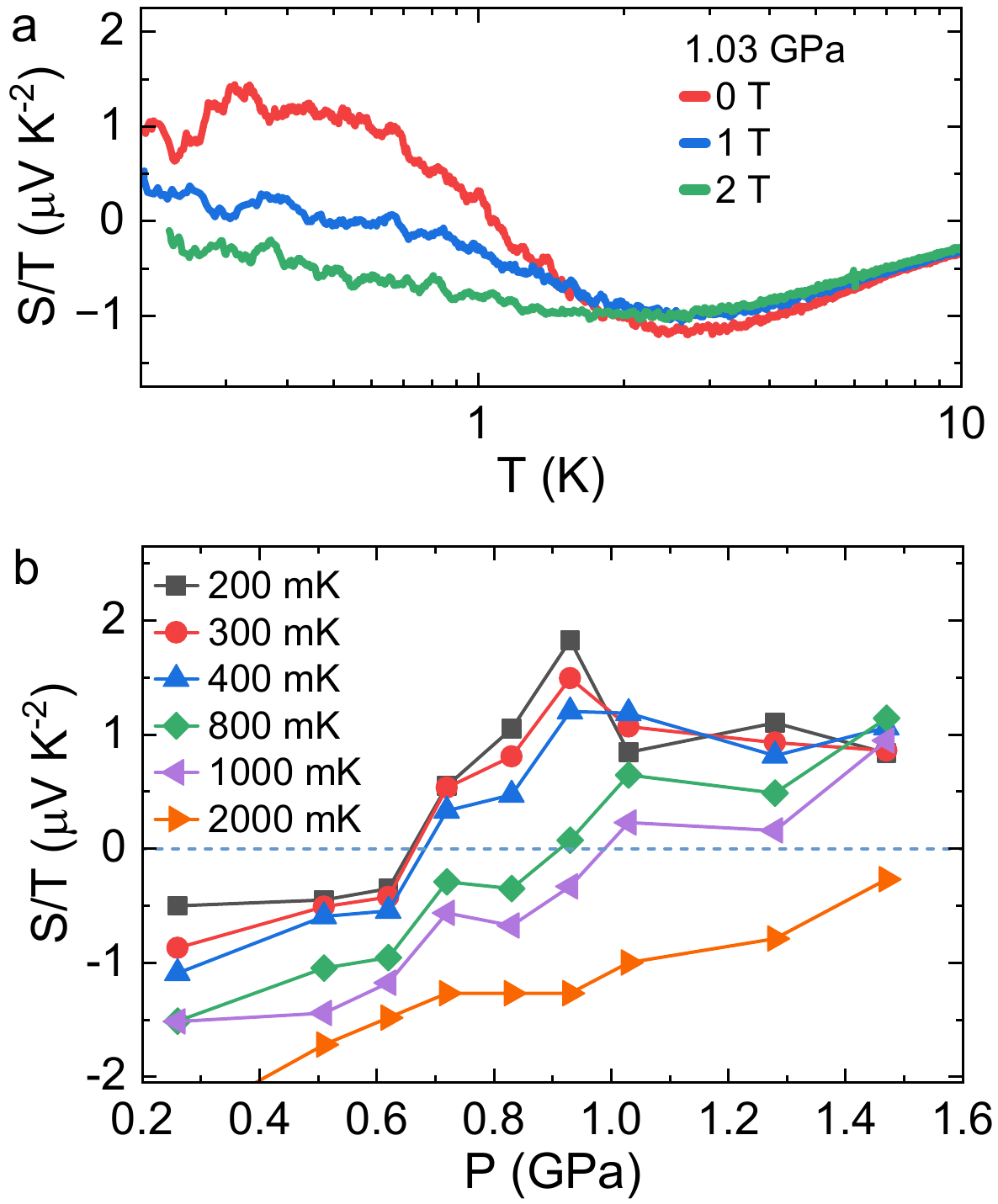}
		\caption{
			(a)~Thermopower divided by temperature versus temperature at 1.03~GPa for the indicated magnetic fields.
			Field was applied parallel to the $c$-axis.
			(b)~Thermopower divided by temperature at the indicated temperatures versus pressure. The dashed blue line corresponds to a value of zero.
		}\label{fig:fig3}
	\end{center}
\end{figure}

At pressures below 0.6 GPa, the pressure-induced increase in orbital-selective hybridization is insufficient to produce a notable response in thermopower, but a clear indication for a Fermi-surface change begins in the pressure range 0.6-0.7 GPa as 4$f$ degrees-of-freedom become more delocalized and further entangled with $c$-states. The continued increase in $f-c$ mixing with increasing pressure terminates long-range order at P$_c$ where there is  quantum criticality of itinerant ferromagnetism allowed by strong spin-orbit coupling and noncentrosymmetry of CeRh$_6$Ge$_4$~\cite{Kirkpatrick2020}.
 A key point is whether the spin-orbit coupling is sufficient, \textit{i.e.}, whether the ratio $E_{SO}/k_BT_F$ is of order one or larger,~\cite{Kirkpatrick2020} where $E_{SO}$ is the spin-orbit splitting and $T_F$ is the Fermi temperature of the renormalized heavy-mass bands.
At ambient pressure, the spin-orbit splitting in CeRh$_6$Ge$_4$ is estimated to be of order 50~meV from a combination of de~Haas-van~Alphen oscillations and DFT calculations~\cite{Wang2021}.
Assuming that $T_F$ is given approximately by the neutron quasi-elastic linewidth (3.3 meV) or by the Kondo scale (about 4.5 meV) estimated from the Sommerfeld coefficient of CeRh$_6$Ge$_4$ with an effectively 4-fold degenerate ground state~\cite{Desgranges2014}, then $E_{SO}/k_{B}T_{F}$ is on the order of 10, and this condition for quantum criticality of itinerant ferromagnetism is satisfied.

The picture of criticality that comes from thermopower measurements is that there are two sequential effects: the first just below 0.7~GPa is driven by orbital-selective hybridization that produces a change in the Fermi surface within the ferromagnetically ordered state and this is followed by a quantum-critical point near P$_c$ that is allowed by strong spin-orbit coupling and a noncentrosymmetric crystal structure.
This picture is at odds with the initially proposed Kondo-breakdown scenario of quantum criticality~\cite{Shen2020} but is consistent with experimental observations reported there.
Reexamination of the relationship between quantum-oscillation measurements and more realistic band calculations that especially include the consequences of strong hybridization would be worthwhile.
The pressure variation of $S/T$ does not follow that of $C/T$ which peaks sharply at P$_c$ and is expected at a QCP.
This is not surprising given that the value of q from Eq.~(1) deviates strongly from unity below and above P$_c$.
As discussed, there are several possible reasons for this.
In the Boltzmann model, the thermopower is determined by $d\ln{\sigma(\epsilon)}/d\epsilon$, where $\sigma$($\epsilon$) is the energy-dependent electrical conductivity at the Fermi energy.
The extreme sensitivity of $S$ to details of the Fermi surface is a strength of thermopower measurements but also its downfall in being able to interpret straightforwardly.
The simple relationship between $S$ and $C$ reflected in Eq.~(1), though instructive, is valid only if the system can be considered a free electron gas, which is not the case in the vicinity of a QCP.
It, therefore, is reasonable that $S/T$ and $C/T$ do not show the same functional dependence on pressure as the QCP is approached in CeRh$_6$Ge$_4$.

\begin{acknowledgments}
Work was primarily supported by the U.S. Department of Energy, Office of Basic Energy Sciences, Division of Materials Science and Engineering project ``Quantum Fluctuations in Narrow-Band Systems''.
S.~Seo and T.~Asaba acknowledge Director’s Fellowships supported by the Los Alamos Laboratory Directed Research and Development program.
Work by S.~Seo in South Korea was supported by Basic Science Research Program through the National Research Foundation of Korea (NRF) funded by the Ministry of Education (NRF-2022R1I1A1A01072925).
We thank M. A. Continentino and T. R. Kirkpatrick for helpful discussions.
\end{acknowledgments}

\bibliography{lib}

\begin{thebibliography}{26}%
\makeatletter
\providecommand \@ifxundefined [1]{%
 \@ifx{#1\undefined}
}%
\providecommand \@ifnum [1]{%
 \ifnum #1\expandafter \@firstoftwo
 \else \expandafter \@secondoftwo
 \fi
}%
\providecommand \@ifx [1]{%
 \ifx #1\expandafter \@firstoftwo
 \else \expandafter \@secondoftwo
 \fi
}%
\providecommand \natexlab [1]{#1}%
\providecommand \enquote  [1]{``#1''}%
\providecommand \bibnamefont  [1]{#1}%
\providecommand \bibfnamefont [1]{#1}%
\providecommand \citenamefont [1]{#1}%
\providecommand \href@noop [0]{\@secondoftwo}%
\providecommand \href [0]{\begingroup \@sanitize@url \@href}%
\providecommand \@href[1]{\@@startlink{#1}\@@href}%
\providecommand \@@href[1]{\endgroup#1\@@endlink}%
\providecommand \@sanitize@url [0]{\catcode `\\12\catcode `\$12\catcode
  `\&12\catcode `\#12\catcode `\^12\catcode `\_12\catcode `\%12\relax}%
\providecommand \@@startlink[1]{}%
\providecommand \@@endlink[0]{}%
\providecommand \url  [0]{\begingroup\@sanitize@url \@url }%
\providecommand \@url [1]{\endgroup\@href {#1}{\urlprefix }}%
\providecommand \urlprefix  [0]{URL }%
\providecommand \Eprint [0]{\href }%
\providecommand \doibase [0]{http://dx.doi.org/}%
\providecommand \selectlanguage [0]{\@gobble}%
\providecommand \bibinfo  [0]{\@secondoftwo}%
\providecommand \bibfield  [0]{\@secondoftwo}%
\providecommand \translation [1]{[#1]}%
\providecommand \BibitemOpen [0]{}%
\providecommand \bibitemStop [0]{}%
\providecommand \bibitemNoStop [0]{.\EOS\space}%
\providecommand \EOS [0]{\spacefactor3000\relax}%
\providecommand \BibitemShut  [1]{\csname bibitem#1\endcsname}%
\let\auto@bib@innerbib\@empty
\bibitem [{\citenamefont {Brando}\ \emph {et~al.}(2016)\citenamefont {Brando},
  \citenamefont {Belitz}, \citenamefont {Grosche},\ and\ \citenamefont
  {Kirkpatrick}}]{Brando2016}%
  \BibitemOpen
  \bibfield  {author} {\bibinfo {author} {\bibfnamefont {M.}~\bibnamefont
  {Brando}}, \bibinfo {author} {\bibfnamefont {D.}~\bibnamefont {Belitz}},
  \bibinfo {author} {\bibfnamefont {F.~M.}\ \bibnamefont {Grosche}}, \ and\
  \bibinfo {author} {\bibfnamefont {T.~R.}\ \bibnamefont {Kirkpatrick}},\
  }\href {\doibase 10.1103/RevModPhys.88.025006} {\bibfield  {journal}
  {\bibinfo  {journal} {Reviews of Modern Physics}\ }\textbf {\bibinfo {volume}
  {88}},\ \bibinfo {pages} {025006} (\bibinfo {year} {2016})},\ \Eprint
  {http://arxiv.org/abs/1502.02898} {arXiv:1502.02898} \BibitemShut {NoStop}%
\bibitem [{\citenamefont {Chen}\ \emph {et~al.}(2022)\citenamefont {Chen},
  \citenamefont {Wang}, \citenamefont {Hu},\ and\ \citenamefont
  {Yang}}]{Chen2022}%
  \BibitemOpen
  \bibfield  {author} {\bibinfo {author} {\bibfnamefont {J.}~\bibnamefont
  {Chen}}, \bibinfo {author} {\bibfnamefont {J.}~\bibnamefont {Wang}}, \bibinfo
  {author} {\bibfnamefont {D.}~\bibnamefont {Hu}}, \ and\ \bibinfo {author}
  {\bibfnamefont {Y.-f.}\ \bibnamefont {Yang}},\ }\href {\doibase
  10.1103/PhysRevB.106.075114} {\bibfield  {journal} {\bibinfo  {journal}
  {Physical Review B}\ }\textbf {\bibinfo {volume} {106}},\ \bibinfo {pages}
  {075114} (\bibinfo {year} {2022})},\ \Eprint
  {http://arxiv.org/abs/2203.15475} {arXiv:2203.15475} \BibitemShut {NoStop}%
\bibitem [{\citenamefont {Kirkpatrick}\ and\ \citenamefont
  {Belitz}(2020)}]{Kirkpatrick2020}%
  \BibitemOpen
  \bibfield  {author} {\bibinfo {author} {\bibfnamefont {T.~R.}\ \bibnamefont
  {Kirkpatrick}}\ and\ \bibinfo {author} {\bibfnamefont {D.}~\bibnamefont
  {Belitz}},\ }\href {\doibase 10.1103/PhysRevLett.124.147201} {\bibfield
  {journal} {\bibinfo  {journal} {Phys. Rev. Lett.}\ }\textbf {\bibinfo
  {volume} {124}},\ \bibinfo {pages} {147201} (\bibinfo {year} {2020})},\
  \Eprint {http://arxiv.org/abs/1911.02649} {arXiv:1911.02649} \BibitemShut
  {NoStop}%
\bibitem [{\citenamefont {Shen}\ \emph {et~al.}(2020)\citenamefont {Shen},
  \citenamefont {Zhang}, \citenamefont {Komijani}, \citenamefont {Nicklas},
  \citenamefont {Borth}, \citenamefont {Wang}, \citenamefont {Chen},
  \citenamefont {Nie}, \citenamefont {Li}, \citenamefont {Lu}, \citenamefont
  {Lee}, \citenamefont {Smidman}, \citenamefont {Steglich}, \citenamefont
  {Coleman},\ and\ \citenamefont {Yuan}}]{Shen2020}%
  \BibitemOpen
  \bibfield  {author} {\bibinfo {author} {\bibfnamefont {B.}~\bibnamefont
  {Shen}}, \bibinfo {author} {\bibfnamefont {Y.}~\bibnamefont {Zhang}},
  \bibinfo {author} {\bibfnamefont {Y.}~\bibnamefont {Komijani}}, \bibinfo
  {author} {\bibfnamefont {M.}~\bibnamefont {Nicklas}}, \bibinfo {author}
  {\bibfnamefont {R.}~\bibnamefont {Borth}}, \bibinfo {author} {\bibfnamefont
  {A.}~\bibnamefont {Wang}}, \bibinfo {author} {\bibfnamefont {Y.}~\bibnamefont
  {Chen}}, \bibinfo {author} {\bibfnamefont {Z.}~\bibnamefont {Nie}}, \bibinfo
  {author} {\bibfnamefont {R.}~\bibnamefont {Li}}, \bibinfo {author}
  {\bibfnamefont {X.}~\bibnamefont {Lu}}, \bibinfo {author} {\bibfnamefont
  {H.}~\bibnamefont {Lee}}, \bibinfo {author} {\bibfnamefont {M.}~\bibnamefont
  {Smidman}}, \bibinfo {author} {\bibfnamefont {F.}~\bibnamefont {Steglich}},
  \bibinfo {author} {\bibfnamefont {P.}~\bibnamefont {Coleman}}, \ and\
  \bibinfo {author} {\bibfnamefont {H.}~\bibnamefont {Yuan}},\ }\href {\doibase
  10.1038/s41586-020-2052-z} {\bibfield  {journal} {\bibinfo  {journal}
  {Nature}\ }\textbf {\bibinfo {volume} {579}},\ \bibinfo {pages} {51}
  (\bibinfo {year} {2020})},\ \Eprint {http://arxiv.org/abs/1907.10470}
  {arXiv:1907.10470} \BibitemShut {NoStop}%
\bibitem [{\citenamefont {Komijani}\ and\ \citenamefont
  {Coleman}(2018)}]{Komijani2018}%
  \BibitemOpen
  \bibfield  {author} {\bibinfo {author} {\bibfnamefont {Y.}~\bibnamefont
  {Komijani}}\ and\ \bibinfo {author} {\bibfnamefont {P.}~\bibnamefont
  {Coleman}},\ }\href {\doibase 10.1103/PhysRevLett.120.157206} {\bibfield
  {journal} {\bibinfo  {journal} {Physical Review Letters}\ }\textbf {\bibinfo
  {volume} {120}},\ \bibinfo {pages} {157206} (\bibinfo {year} {2018})},\
  \Eprint {http://arxiv.org/abs/1710.03345} {arXiv:1710.03345} \BibitemShut
  {NoStop}%
\bibitem [{\citenamefont {Kotegawa}\ \emph {et~al.}(2019)\citenamefont
  {Kotegawa}, \citenamefont {Matsuoka}, \citenamefont {Uga}, \citenamefont
  {Takemura}, \citenamefont {Manago}, \citenamefont {Chikuchi}, \citenamefont
  {Sugawara}, \citenamefont {Tou},\ and\ \citenamefont
  {Harima}}]{Kotegawa2019}%
  \BibitemOpen
  \bibfield  {author} {\bibinfo {author} {\bibfnamefont {H.}~\bibnamefont
  {Kotegawa}}, \bibinfo {author} {\bibfnamefont {E.}~\bibnamefont {Matsuoka}},
  \bibinfo {author} {\bibfnamefont {T.}~\bibnamefont {Uga}}, \bibinfo {author}
  {\bibfnamefont {M.}~\bibnamefont {Takemura}}, \bibinfo {author}
  {\bibfnamefont {M.}~\bibnamefont {Manago}}, \bibinfo {author} {\bibfnamefont
  {N.}~\bibnamefont {Chikuchi}}, \bibinfo {author} {\bibfnamefont
  {H.}~\bibnamefont {Sugawara}}, \bibinfo {author} {\bibfnamefont
  {H.}~\bibnamefont {Tou}}, \ and\ \bibinfo {author} {\bibfnamefont
  {H.}~\bibnamefont {Harima}},\ }\href {\doibase 10.7566/JPSJ.88.093702}
  {\bibfield  {journal} {\bibinfo  {journal} {Journal of the Physical Society
  of Japan}\ }\textbf {\bibinfo {volume} {88}},\ \bibinfo {pages} {093702}
  (\bibinfo {year} {2019})},\ \Eprint {http://arxiv.org/abs/1907.09802}
  {arXiv:1907.09802} \BibitemShut {NoStop}%
\bibitem [{\citenamefont {Wang}\ \emph {et~al.}(2021)\citenamefont {Wang},
  \citenamefont {Du}, \citenamefont {Zhang}, \citenamefont {Graf},
  \citenamefont {Shen}, \citenamefont {Chen}, \citenamefont {Liu},
  \citenamefont {Smidman}, \citenamefont {Cao}, \citenamefont {Steglich},\ and\
  \citenamefont {Yuan}}]{Wang2021}%
  \BibitemOpen
  \bibfield  {author} {\bibinfo {author} {\bibfnamefont {A.}~\bibnamefont
  {Wang}}, \bibinfo {author} {\bibfnamefont {F.}~\bibnamefont {Du}}, \bibinfo
  {author} {\bibfnamefont {Y.}~\bibnamefont {Zhang}}, \bibinfo {author}
  {\bibfnamefont {D.}~\bibnamefont {Graf}}, \bibinfo {author} {\bibfnamefont
  {B.}~\bibnamefont {Shen}}, \bibinfo {author} {\bibfnamefont {Y.}~\bibnamefont
  {Chen}}, \bibinfo {author} {\bibfnamefont {Y.}~\bibnamefont {Liu}}, \bibinfo
  {author} {\bibfnamefont {M.}~\bibnamefont {Smidman}}, \bibinfo {author}
  {\bibfnamefont {C.}~\bibnamefont {Cao}}, \bibinfo {author} {\bibfnamefont
  {F.}~\bibnamefont {Steglich}}, \ and\ \bibinfo {author} {\bibfnamefont
  {H.}~\bibnamefont {Yuan}},\ }\href {\doibase 10.1016/j.scib.2021.03.006}
  {\bibfield  {journal} {\bibinfo  {journal} {Science Bulletin}\ }\textbf
  {\bibinfo {volume} {66}},\ \bibinfo {pages} {1389} (\bibinfo {year}
  {2021})},\ \Eprint {http://arxiv.org/abs/2101.08972} {arXiv:2101.08972}
  \BibitemShut {NoStop}%
\bibitem [{\citenamefont {Wu}\ \emph {et~al.}(2021)\citenamefont {Wu},
  \citenamefont {Zhang}, \citenamefont {Du}, \citenamefont {Shen},
  \citenamefont {Zheng}, \citenamefont {Fang}, \citenamefont {Smidman},
  \citenamefont {Cao}, \citenamefont {Steglich}, \citenamefont {Yuan},
  \citenamefont {Denlinger},\ and\ \citenamefont {Liu}}]{Wu2021}%
  \BibitemOpen
  \bibfield  {author} {\bibinfo {author} {\bibfnamefont {Y.}~\bibnamefont
  {Wu}}, \bibinfo {author} {\bibfnamefont {Y.}~\bibnamefont {Zhang}}, \bibinfo
  {author} {\bibfnamefont {F.}~\bibnamefont {Du}}, \bibinfo {author}
  {\bibfnamefont {B.}~\bibnamefont {Shen}}, \bibinfo {author} {\bibfnamefont
  {H.}~\bibnamefont {Zheng}}, \bibinfo {author} {\bibfnamefont
  {Y.}~\bibnamefont {Fang}}, \bibinfo {author} {\bibfnamefont {M.}~\bibnamefont
  {Smidman}}, \bibinfo {author} {\bibfnamefont {C.}~\bibnamefont {Cao}},
  \bibinfo {author} {\bibfnamefont {F.}~\bibnamefont {Steglich}}, \bibinfo
  {author} {\bibfnamefont {H.}~\bibnamefont {Yuan}}, \bibinfo {author}
  {\bibfnamefont {J.~D.}\ \bibnamefont {Denlinger}}, \ and\ \bibinfo {author}
  {\bibfnamefont {Y.}~\bibnamefont {Liu}},\ }\href {\doibase
  10.1103/PhysRevLett.126.216406} {\bibfield  {journal} {\bibinfo  {journal}
  {Physical Review Letters}\ }\textbf {\bibinfo {volume} {126}},\ \bibinfo
  {pages} {216406} (\bibinfo {year} {2021})},\ \Eprint
  {http://arxiv.org/abs/2104.03600} {arXiv:2104.03600} \BibitemShut {NoStop}%
\bibitem [{\citenamefont {Pei}\ \emph {et~al.}(2021)\citenamefont {Pei},
  \citenamefont {Zhang}, \citenamefont {Wei}, \citenamefont {Chen},
  \citenamefont {Hu}, \citenamefont {Yang}, \citenamefont {Yuan},\ and\
  \citenamefont {Qi}}]{Pei2021a}%
  \BibitemOpen
  \bibfield  {author} {\bibinfo {author} {\bibfnamefont {Y.~H.}\ \bibnamefont
  {Pei}}, \bibinfo {author} {\bibfnamefont {Y.~J.}\ \bibnamefont {Zhang}},
  \bibinfo {author} {\bibfnamefont {Z.~X.}\ \bibnamefont {Wei}}, \bibinfo
  {author} {\bibfnamefont {Y.~X.}\ \bibnamefont {Chen}}, \bibinfo {author}
  {\bibfnamefont {K.}~\bibnamefont {Hu}}, \bibinfo {author} {\bibfnamefont
  {Y.-f.}\ \bibnamefont {Yang}}, \bibinfo {author} {\bibfnamefont {H.~Q.}\
  \bibnamefont {Yuan}}, \ and\ \bibinfo {author} {\bibfnamefont
  {J.}~\bibnamefont {Qi}},\ }\href {\doibase 10.1103/PhysRevB.103.L180409}
  {\bibfield  {journal} {\bibinfo  {journal} {Physical Review B}\ }\textbf
  {\bibinfo {volume} {103}},\ \bibinfo {pages} {L180409} (\bibinfo {year}
  {2021})},\ \Eprint {http://arxiv.org/abs/2102.08572} {arXiv:2102.08572}
  \BibitemShut {NoStop}%
\bibitem [{\citenamefont {Shu}\ \emph {et~al.}(2021)\citenamefont {Shu},
  \citenamefont {Adroja}, \citenamefont {Hillier}, \citenamefont {Zhang},
  \citenamefont {Chen}, \citenamefont {Shen}, \citenamefont {Orlandi},
  \citenamefont {Walker}, \citenamefont {Liu}, \citenamefont {Cao},
  \citenamefont {Steglich}, \citenamefont {Yuan},\ and\ \citenamefont
  {Smidman}}]{Shu2021}%
  \BibitemOpen
  \bibfield  {author} {\bibinfo {author} {\bibfnamefont {J.~W.}\ \bibnamefont
  {Shu}}, \bibinfo {author} {\bibfnamefont {D.~T.}\ \bibnamefont {Adroja}},
  \bibinfo {author} {\bibfnamefont {A.~D.}\ \bibnamefont {Hillier}}, \bibinfo
  {author} {\bibfnamefont {Y.~J.}\ \bibnamefont {Zhang}}, \bibinfo {author}
  {\bibfnamefont {Y.~X.}\ \bibnamefont {Chen}}, \bibinfo {author}
  {\bibfnamefont {B.}~\bibnamefont {Shen}}, \bibinfo {author} {\bibfnamefont
  {F.}~\bibnamefont {Orlandi}}, \bibinfo {author} {\bibfnamefont {H.~C.}\
  \bibnamefont {Walker}}, \bibinfo {author} {\bibfnamefont {Y.}~\bibnamefont
  {Liu}}, \bibinfo {author} {\bibfnamefont {C.}~\bibnamefont {Cao}}, \bibinfo
  {author} {\bibfnamefont {F.}~\bibnamefont {Steglich}}, \bibinfo {author}
  {\bibfnamefont {H.~Q.}\ \bibnamefont {Yuan}}, \ and\ \bibinfo {author}
  {\bibfnamefont {M.}~\bibnamefont {Smidman}},\ }\href {\doibase
  10.1103/PhysRevB.104.L140411} {\bibfield  {journal} {\bibinfo  {journal}
  {Physical Review B}\ }\textbf {\bibinfo {volume} {104}},\ \bibinfo {pages}
  {L140411} (\bibinfo {year} {2021})},\ \Eprint
  {http://arxiv.org/abs/2102.12788} {arXiv:2102.12788} \BibitemShut {NoStop}%
\bibitem [{\citenamefont {Yuan}(2023)}]{Huiqiu2023}%
  \BibitemOpen
  \bibfield  {author} {\bibinfo {author} {\bibfnamefont {H.}~\bibnamefont
  {Yuan}}\ }(\bibinfo  {publisher} {Presented at International Conference of
  Strongly Correlated Electron Systems},\ \bibinfo {year} {2023})\BibitemShut
  {NoStop}%
\bibitem [{\citenamefont {Hartmann}\ \emph {et~al.}(2010)\citenamefont
  {Hartmann}, \citenamefont {Oeschler}, \citenamefont {Krellner}, \citenamefont
  {Geibel}, \citenamefont {Paschen},\ and\ \citenamefont
  {Steglich}}]{Hartmann2010}%
  \BibitemOpen
  \bibfield  {author} {\bibinfo {author} {\bibfnamefont {S.}~\bibnamefont
  {Hartmann}}, \bibinfo {author} {\bibfnamefont {N.}~\bibnamefont {Oeschler}},
  \bibinfo {author} {\bibfnamefont {C.}~\bibnamefont {Krellner}}, \bibinfo
  {author} {\bibfnamefont {C.}~\bibnamefont {Geibel}}, \bibinfo {author}
  {\bibfnamefont {S.}~\bibnamefont {Paschen}}, \ and\ \bibinfo {author}
  {\bibfnamefont {F.}~\bibnamefont {Steglich}},\ }\href {\doibase
  10.1103/PhysRevLett.104.096401} {\bibfield  {journal} {\bibinfo  {journal}
  {Physical Review Letters}\ }\textbf {\bibinfo {volume} {104}},\ \bibinfo
  {pages} {096401} (\bibinfo {year} {2010})}\BibitemShut {NoStop}%
\bibitem [{\citenamefont {Luo}\ \emph {et~al.}(2018)\citenamefont {Luo},
  \citenamefont {Lu}, \citenamefont {Dioguardi}, \citenamefont {Rosa},
  \citenamefont {Bauer}, \citenamefont {Si},\ and\ \citenamefont
  {Thompson}}]{Luo2018}%
  \BibitemOpen
  \bibfield  {author} {\bibinfo {author} {\bibfnamefont {Y.}~\bibnamefont
  {Luo}}, \bibinfo {author} {\bibfnamefont {X.}~\bibnamefont {Lu}}, \bibinfo
  {author} {\bibfnamefont {A.~P.}\ \bibnamefont {Dioguardi}}, \bibinfo {author}
  {\bibfnamefont {P.~S.~F.}\ \bibnamefont {Rosa}}, \bibinfo {author}
  {\bibfnamefont {E.~D.}\ \bibnamefont {Bauer}}, \bibinfo {author}
  {\bibfnamefont {Q.}~\bibnamefont {Si}}, \ and\ \bibinfo {author}
  {\bibfnamefont {J.~D.}\ \bibnamefont {Thompson}},\ }\href {\doibase
  10.1038/s41535-018-0080-9} {\bibfield  {journal} {\bibinfo  {journal} {npj
  Quantum Materials}\ }\textbf {\bibinfo {volume} {3}},\ \bibinfo {pages} {6}
  (\bibinfo {year} {2018})},\ \Eprint {http://arxiv.org/abs/1606.07848}
  {arXiv:1606.07848} \BibitemShut {NoStop}%
\bibitem [{\citenamefont {Vo{\ss}winkel}\ \emph {et~al.}(2012)\citenamefont
  {Vo{\ss}winkel}, \citenamefont {Niehaus}, \citenamefont {Rodewald},\ and\
  \citenamefont {P{\"{o}}ttgen}}]{Vosswinkel}%
  \BibitemOpen
  \bibfield  {author} {\bibinfo {author} {\bibfnamefont {D.}~\bibnamefont
  {Vo{\ss}winkel}}, \bibinfo {author} {\bibfnamefont {O.}~\bibnamefont
  {Niehaus}}, \bibinfo {author} {\bibfnamefont {U.~C.}\ \bibnamefont
  {Rodewald}}, \ and\ \bibinfo {author} {\bibfnamefont {R.}~\bibnamefont
  {P{\"{o}}ttgen}},\ }\href {\doibase 10.5560/znb.2012-0265} {\bibfield
  {journal} {\bibinfo  {journal} {Zeitschrift f{\"{u}}r Naturforschung B}\
  }\textbf {\bibinfo {volume} {67}},\ \bibinfo {pages} {1241} (\bibinfo {year}
  {2012})}\BibitemShut {NoStop}%
\bibitem [{\citenamefont {Luo}\ \emph {et~al.}(2016)\citenamefont {Luo},
  \citenamefont {Rosa}, \citenamefont {Bauer},\ and\ \citenamefont
  {Thompson}}]{Luo2016}%
  \BibitemOpen
  \bibfield  {author} {\bibinfo {author} {\bibfnamefont {Y.}~\bibnamefont
  {Luo}}, \bibinfo {author} {\bibfnamefont {P.~F.~S.}\ \bibnamefont {Rosa}},
  \bibinfo {author} {\bibfnamefont {E.~D.}\ \bibnamefont {Bauer}}, \ and\
  \bibinfo {author} {\bibfnamefont {J.~D.}\ \bibnamefont {Thompson}},\ }\href
  {\doibase 10.1103/PhysRevB.93.201102} {\bibfield  {journal} {\bibinfo
  {journal} {Physical Review B}\ }\textbf {\bibinfo {volume} {93}},\ \bibinfo
  {pages} {201102} (\bibinfo {year} {2016})},\ \Eprint
  {http://arxiv.org/abs/1602.08069} {arXiv:1602.08069} \BibitemShut {NoStop}%
\bibitem [{\citenamefont {Stockert}\ and\ \citenamefont
  {Oeschler}(2011)}]{Stockert2011}%
  \BibitemOpen
  \bibfield  {author} {\bibinfo {author} {\bibfnamefont {U.}~\bibnamefont
  {Stockert}}\ and\ \bibinfo {author} {\bibfnamefont {N.}~\bibnamefont
  {Oeschler}},\ }\href {\doibase 10.1016/j.cryogenics.2010.12.009} {\bibfield
  {journal} {\bibinfo  {journal} {Cryogenics}\ }\textbf {\bibinfo {volume}
  {51}},\ \bibinfo {pages} {154} (\bibinfo {year} {2011})}\BibitemShut
  {NoStop}%
\bibitem [{\citenamefont {Miyake}\ and\ \citenamefont
  {Narikiyo}(2002)}]{Miyake2002}%
  \BibitemOpen
  \bibfield  {author} {\bibinfo {author} {\bibfnamefont {K.}~\bibnamefont
  {Miyake}}\ and\ \bibinfo {author} {\bibfnamefont {O.}~\bibnamefont
  {Narikiyo}},\ }\href {\doibase 10.1143/JPSJ.71.867} {\bibfield  {journal}
  {\bibinfo  {journal} {Journal of the Physical Society of Japan}\ }\textbf
  {\bibinfo {volume} {71}},\ \bibinfo {pages} {867} (\bibinfo {year}
  {2002})}\BibitemShut {NoStop}%
\bibitem [{\citenamefont {Behnia}\ \emph {et~al.}(2004)\citenamefont {Behnia},
  \citenamefont {Jaccard},\ and\ \citenamefont {Flouquet}}]{Behnia2004}%
  \BibitemOpen
  \bibfield  {author} {\bibinfo {author} {\bibfnamefont {K.}~\bibnamefont
  {Behnia}}, \bibinfo {author} {\bibfnamefont {D.}~\bibnamefont {Jaccard}}, \
  and\ \bibinfo {author} {\bibfnamefont {J.}~\bibnamefont {Flouquet}},\ }\href
  {\doibase 10.1088/0953-8984/16/28/037} {\bibfield  {journal} {\bibinfo
  {journal} {Journal of Physics: Condensed Matter}\ }\textbf {\bibinfo {volume}
  {16}},\ \bibinfo {pages} {5187} (\bibinfo {year} {2004})},\ \Eprint
  {http://arxiv.org/abs/0405030} {arXiv:0405030 [cond-mat]} \BibitemShut
  {NoStop}%
\bibitem [{\citenamefont {Vojta}(2010)}]{Vojta2010}%
  \BibitemOpen
  \bibfield  {author} {\bibinfo {author} {\bibfnamefont {M.}~\bibnamefont
  {Vojta}},\ }\href {\doibase 10.1007/s10909-010-0206-3} {\bibfield  {journal}
  {\bibinfo  {journal} {Journal of Low Temperature Physics}\ }\textbf {\bibinfo
  {volume} {161}},\ \bibinfo {pages} {203} (\bibinfo {year} {2010})},\ \Eprint
  {http://arxiv.org/abs/1006.1559} {arXiv:1006.1559} \BibitemShut {NoStop}%
\bibitem [{\citenamefont {Kim}\ and\ \citenamefont
  {P{\'{e}}pin}(2010)}]{Kim2010}%
  \BibitemOpen
  \bibfield  {author} {\bibinfo {author} {\bibfnamefont {K.-S.}\ \bibnamefont
  {Kim}}\ and\ \bibinfo {author} {\bibfnamefont {C.}~\bibnamefont
  {P{\'{e}}pin}},\ }\href {\doibase 10.1103/PhysRevB.81.205108} {\bibfield
  {journal} {\bibinfo  {journal} {Physical Review B}\ }\textbf {\bibinfo
  {volume} {81}},\ \bibinfo {pages} {205108} (\bibinfo {year} {2010})},\
  \Eprint {http://arxiv.org/abs/1002.2612} {arXiv:1002.2612} \BibitemShut
  {NoStop}%
\bibitem [{\citenamefont {Matsuoka}\ \emph {et~al.}(2015)\citenamefont
  {Matsuoka}, \citenamefont {Hondo}, \citenamefont {Fujii}, \citenamefont
  {Oshima}, \citenamefont {Sugawara}, \citenamefont {Sakurai}, \citenamefont
  {Ohta}, \citenamefont {Kneidinger}, \citenamefont {Salamakha}, \citenamefont
  {Michor},\ and\ \citenamefont {Bauer}}]{Matsuoka2015}%
  \BibitemOpen
  \bibfield  {author} {\bibinfo {author} {\bibfnamefont {E.}~\bibnamefont
  {Matsuoka}}, \bibinfo {author} {\bibfnamefont {C.}~\bibnamefont {Hondo}},
  \bibinfo {author} {\bibfnamefont {T.}~\bibnamefont {Fujii}}, \bibinfo
  {author} {\bibfnamefont {A.}~\bibnamefont {Oshima}}, \bibinfo {author}
  {\bibfnamefont {H.}~\bibnamefont {Sugawara}}, \bibinfo {author}
  {\bibfnamefont {T.}~\bibnamefont {Sakurai}}, \bibinfo {author} {\bibfnamefont
  {H.}~\bibnamefont {Ohta}}, \bibinfo {author} {\bibfnamefont {F.}~\bibnamefont
  {Kneidinger}}, \bibinfo {author} {\bibfnamefont {L.}~\bibnamefont
  {Salamakha}}, \bibinfo {author} {\bibfnamefont {H.}~\bibnamefont {Michor}}, \
  and\ \bibinfo {author} {\bibfnamefont {E.}~\bibnamefont {Bauer}},\ }\href
  {\doibase 10.7566/JPSJ.84.073704} {\bibfield  {journal} {\bibinfo  {journal}
  {Journal of the Physical Society of Japan}\ }\textbf {\bibinfo {volume}
  {84}},\ \bibinfo {pages} {073704} (\bibinfo {year} {2015})}\BibitemShut
  {NoStop}%
\bibitem [{\citenamefont {Miyake}\ and\ \citenamefont
  {Kohno}(2005)}]{Miyake2005}%
  \BibitemOpen
  \bibfield  {author} {\bibinfo {author} {\bibfnamefont {K.}~\bibnamefont
  {Miyake}}\ and\ \bibinfo {author} {\bibfnamefont {H.}~\bibnamefont {Kohno}},\
  }\href {\doibase 10.1143/JPSJ.74.254} {\bibfield  {journal} {\bibinfo
  {journal} {Journal of the Physical Society of Japan}\ }\textbf {\bibinfo
  {volume} {74}},\ \bibinfo {pages} {254} (\bibinfo {year} {2005})}\BibitemShut
  {NoStop}%
\bibitem [{\citenamefont {Liu}\ \emph {et~al.}(2023)\citenamefont {Liu},
  \citenamefont {Paschen},\ and\ \citenamefont {Si}}]{Liu2023a}%
  \BibitemOpen
  \bibfield  {author} {\bibinfo {author} {\bibfnamefont {C.-C.}\ \bibnamefont
  {Liu}}, \bibinfo {author} {\bibfnamefont {S.}~\bibnamefont {Paschen}}, \ and\
  \bibinfo {author} {\bibfnamefont {Q.}~\bibnamefont {Si}},\ }\href {\doibase
  10.1073/pnas.2300903120} {\bibfield  {journal} {\bibinfo  {journal}
  {Proceedings of the National Academy of Sciences}\ }\textbf {\bibinfo
  {volume} {120}},\ \bibinfo {pages} {e2300903120} (\bibinfo {year}
  {2023})}\BibitemShut {NoStop}%
\bibitem [{\citenamefont {Martelli}\ \emph {et~al.}(2019)\citenamefont
  {Martelli}, \citenamefont {Cai}, \citenamefont {Nica}, \citenamefont
  {Taupin}, \citenamefont {Prokofiev}, \citenamefont {Liu}, \citenamefont
  {Lai}, \citenamefont {Yu}, \citenamefont {Ingersent}, \citenamefont
  {K{\"{u}}chler}, \citenamefont {Strydom}, \citenamefont {Geiger},
  \citenamefont {Haenel}, \citenamefont {Larrea}, \citenamefont {Si},\ and\
  \citenamefont {Paschen}}]{Martelli2019}%
  \BibitemOpen
  \bibfield  {author} {\bibinfo {author} {\bibfnamefont {V.}~\bibnamefont
  {Martelli}}, \bibinfo {author} {\bibfnamefont {A.}~\bibnamefont {Cai}},
  \bibinfo {author} {\bibfnamefont {E.~M.}\ \bibnamefont {Nica}}, \bibinfo
  {author} {\bibfnamefont {M.}~\bibnamefont {Taupin}}, \bibinfo {author}
  {\bibfnamefont {A.}~\bibnamefont {Prokofiev}}, \bibinfo {author}
  {\bibfnamefont {C.-C.}\ \bibnamefont {Liu}}, \bibinfo {author} {\bibfnamefont
  {H.-H.}\ \bibnamefont {Lai}}, \bibinfo {author} {\bibfnamefont
  {R.}~\bibnamefont {Yu}}, \bibinfo {author} {\bibfnamefont {K.}~\bibnamefont
  {Ingersent}}, \bibinfo {author} {\bibfnamefont {R.}~\bibnamefont
  {K{\"{u}}chler}}, \bibinfo {author} {\bibfnamefont {A.~M.}\ \bibnamefont
  {Strydom}}, \bibinfo {author} {\bibfnamefont {D.}~\bibnamefont {Geiger}},
  \bibinfo {author} {\bibfnamefont {J.}~\bibnamefont {Haenel}}, \bibinfo
  {author} {\bibfnamefont {J.}~\bibnamefont {Larrea}}, \bibinfo {author}
  {\bibfnamefont {Q.}~\bibnamefont {Si}}, \ and\ \bibinfo {author}
  {\bibfnamefont {S.}~\bibnamefont {Paschen}},\ }\href {\doibase
  10.1073/pnas.1908101116} {\bibfield  {journal} {\bibinfo  {journal}
  {Proceedings of the National Academy of Sciences}\ }\textbf {\bibinfo
  {volume} {116}},\ \bibinfo {pages} {17701} (\bibinfo {year}
  {2019})}\BibitemShut {NoStop}%
\bibitem [{\citenamefont {Thompson}\ and\ \citenamefont
  {Lawrence}(1994)}]{Thompson1994}%
  \BibitemOpen
  \bibfield  {author} {\bibinfo {author} {\bibfnamefont {J.}~\bibnamefont
  {Thompson}}\ and\ \bibinfo {author} {\bibfnamefont {J.}~\bibnamefont
  {Lawrence}},\ }in\ \href {\doibase 10.1016/S0168-1273(05)80062-5} {\emph
  {\bibinfo {booktitle} {Handbook on the Physics and Chemistry of Rare
  Earths}}},\ Vol.~\bibinfo {volume} {19}\ (\bibinfo {year} {1994})\ pp.\
  \bibinfo {pages} {383--478}\BibitemShut {NoStop}%
\bibitem [{\citenamefont {Desgranges}(2014)}]{Desgranges2014}%
  \BibitemOpen
  \bibfield  {author} {\bibinfo {author} {\bibfnamefont {H.-U.}\ \bibnamefont
  {Desgranges}},\ }\href {\doibase 10.1016/j.physb.2014.07.077} {\bibfield
  {journal} {\bibinfo  {journal} {Physica B: Condensed Matter}\ }\textbf
  {\bibinfo {volume} {454}},\ \bibinfo {pages} {135} (\bibinfo {year}
  {2014})}\BibitemShut {NoStop}%
\end{thebibliography}%

\end{document}